\documentclass[]{emulateapj}
\usepackage{latexsym,natbib}
\usepackage{graphicx}

\begin{document}
\title{The identification of MAXI J1659--152 as a black hole candidate}
\author{M. Kalamkar\altaffilmark{1}, J. Homan\altaffilmark{2}, D. Altamirano\altaffilmark{1}, M. van der Klis\altaffilmark{1}, P. Casella\altaffilmark{3},  M. Linares\altaffilmark{2}}
\altaffiltext{1}{Astronomical Institute, ``Anton Pannekoek'', University of Amsterdam, Science Park 904, 1098 XH, Amsterdam, The Netherlands}
\altaffiltext{2}{MIT Kavli Institute for Astrophysics and Space Research, 70 Vassar Street, Cambridge, MA 02139, USA}
\altaffiltext{3}{School of Physics and Astronomy, University of Southampton, Southampton, Hampshire SO17 1BJ}

\email{m.n.kalamkar@uva.nl}

\begin{abstract}
  We report on the analysis of all 65 pointed {\it Rossi X-ray Timing
    Explorer} observations of the recently discovered soft X-ray
  transient MAXI J1659--152 (initially referred to as GRB
  100925A). The source was studied in terms of its evolution through
  the hardness-intensity diagram (HID) as well as its X-ray
  variability properties. MAXI J1659--152 traced out an anti-clockwise
  loop in the HID, which is commonly seen in transient low-mass X-ray
  binaries. The variability properties of the source, in particular
  the detection of type-B and type-C low-frequency quasi-periodic
  oscillations, and the way they evolve along the HID track, indicate
  that MAXI J1659--152 is a black hole candidate. The spectral and
  variability properties of MAXI J1659--152 imply that the source was
  observed in the hard and soft intermediate states during the {\it
    RXTE} observations, with several transitions between these two
  states.

\end{abstract}

\section{Introduction}\label{intro}
Black hole X-ray binaries have been studied since the early 1970s. It
is now well established that the X-ray spectral and variability
properties of these sources are strongly correlated,
which is most clearly seen in the transient black hole X-ray binaries
(BHTs).  While there is a great variety in the observed outburst
behavior of BHTs (even for single sources), their outbursts typically
proceed along 'q'-shaped tracks in hardness-intensity diagrams (HIDs; 
see, e.g., \citealt{Homan05} and \citealt{Dunn10}), which are traced
out in an anti-clockwise manner; we note that similar tracks are also
traced out by neutron star transients
\citep[e.g.][]{Tudose09,Linares09}. The various branches of these
tracks correspond to distinct spectral states. Not all sources show
all possible states and the time they spend in each state can differ
significantly.

There are various conventions for describing the spectral (and
variability) states of BHTs; see, e.g., reviews by
\citet{Homan05}, \citet{Remillard06}. In this paper we follow the convention used in \citet{Belloni10b}, which is based on the
work by \citet{Belloni05} on GX 339-4.  This source is often used as a
template for the outburst evolution of BHTs, owing to its well defined
q-shaped HID tracks, and the fact that it shows behavior that is common
to many other systems.

When GX 339-4 goes into outburst, its intensity is low and spectrum is
hard (low-hard state or LHS). As the intensity increases, the spectrum
remains hard, until the source makes a transition to the intermediate
state (IMS), where the hard color starts to decrease at a rather
constant intensity. The IMS can be divided into a hard and a soft IMS
(HIMS and SIMS, respectively) depending on the spectral and
variability characteristics observed. The transition from the LHS is
always first to the HIMS. GX 339-4 often shows several transitions
between HIMS and SIMS until the hardness decreases even further and
the source reaches the so called high-soft state (HSS),
where subsequently the hardness remains approximately constant as the
intensity eventually decreases. At some point during this decrease
hardness increases again and the source transits from the HSS via the
IMS back to the LHS, and returns to quiescence.

A BHT exhibits various types of quasi-periodic oscillations
(QPOs) and broad-band noise components, whose properties are strongly
correlated with spectral state \citep[see, e.g.,][]{Klein07}. The
power spectrum of the LHS is characterized by strong broadband
variability (0.01-100 Hz fractional rms amplitude up to 50\%). During
the HIMS so-called type-C QPOs are observed \citep[see][for QPO type
definitions]{Wijnands99,Remillard02,Casella05}, often with strong
harmonic content and accompanied by strong broadband variability
(fractional rms up to 30\%). In the SIMS various types of variability
are observed: power spectra with type-B QPOs (also with strong
harmonic content) or (weaker) type-A QPOs, but also power spectra with
weak peaked noise and/or QPO features that have been poorly
characterized. All SIMS power spectra have in common that their
broadband variability is (considerably) weaker than in the LHS and
HIMS. Type-C QPOs are typically observed between 0.1--10 Hz, whereas
type A and B QPOs are generally confined to the 4--8 Hz range. In the
HSS variability reaches its minimum strength (a few percent rms);
sometimes very weak QPOs above 10 Hz are seen \citep{Homan01}. 

In this paper we present a study of {\it  Rossi X-ray Timing
Explorer  (RXTE)} observations of MAXI J1659--152
(henceforth J1659). This source was discovered with the {\it Swift}
Burst Alert Telescope \citep[BAT;][]{Barthelmy05} on September 25, 2010 and was initially thought to
be a gamma-ray burst \citep[GRB 100925A, see][]{Mangano10}. Later, it was
suggested by \cite{Kann10} to be a new Galactic X-ray transient due to its
persistent X-ray emission; this was confirmed by \cite{Negoro10} with
{\it MAXI}  observations. J1659 has also been detected in the radio
\citep{Vanderhorst10}, optical \citep{Jelinek10} and sub-mm bands
\citep{deugartepostigo10}. The aim of this letter is to discuss the
results of the aperiodic timing analysis and color evolution from {\it RXTE} observations, based on which we conclude that J1659 is a black hole candidate \citep{Kalamkar10}.

\section{Observations and data analysis}\label{obdata}

We analyzed all 65 {\it RXTE} Proportional Counter Array
\citep[PCA][]{Jahoda06} observations of J1659 taken between September
28, 2010 (MJD 55467) and November 8, 2010 (MJD 55508), after which the source was
not observed due to solar viewing constraints. The intensity and colors were obtained using the Standard
2 data from all the PCUs active during the observations. The
intensity is defined as the count rates in the
2.0--20.0 keV band; hard color as the count rate ratio of the 16.0--20 keV and 2.0--6.0 keV bands. Count rates are background corrected and normalized using the Crab
nebula observations. This method of normalization
\citep{Kuulkers94} is based on the assumption that the Crab nebula is
constant in intensity and hard color. It is used to correct for
differences between PCUs \cite[see, e.g.][for details]{Straaten05}.

\begin{figure}
\center
\resizebox{1\columnwidth}{!}{\rotatebox{-90}{\includegraphics{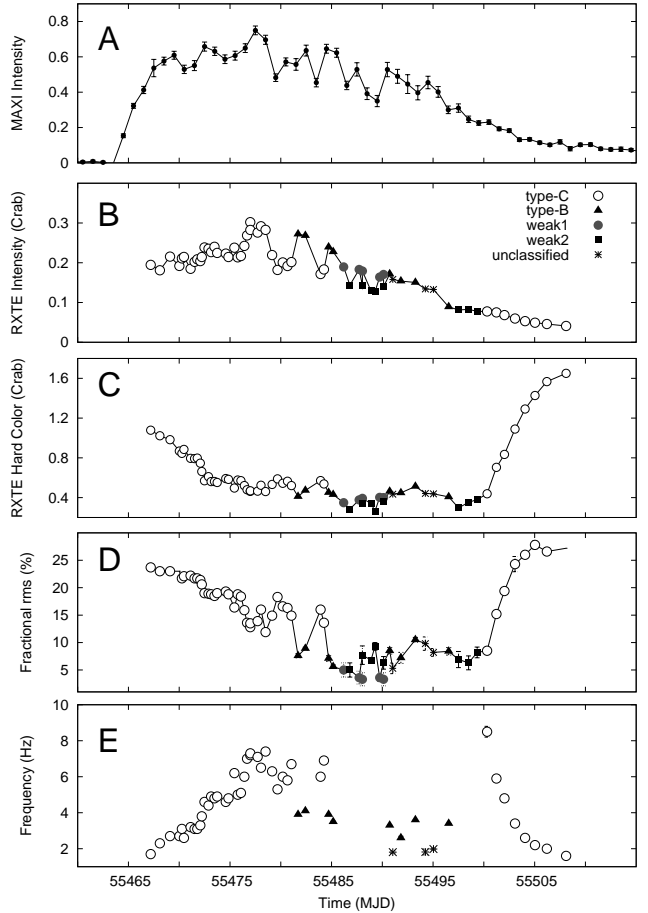}}}
\caption{Top to bottom: From {\it MAXI} observations -- (A) intensity (cnts cm$^{-2}$ s$^{-1}$, 1.5-20.0 keV); from {\it RXTE} observations -- (B)  intensity  (2-20 keV),  (C) hard color (16.0-20.0/2.0-6.0 count rate ratio), (D) average 0.01--100 Hz fractional rms amplitude, and  (E) QPO frequency.  Each point represents one observation, except for the MJD 55488.02 and 55490.11 observations, for which we use two points each. Different  types of power spectra are indicated using different symbols as  follows: (i) open circles: type--C QPOs which indicate the HIMS;  (ii) filled triangles: type--B QPOs; (iii)  filled circles and filled squares are observations with weak  variability (iv) stars are the observations with unclassified power spectra. The latter three categories comprise the SIMS.} \label{fig:inhcrms}
\end{figure} 

The timing analysis was done with the Event mode data in the 2--60 keV
range. Power spectra were generated for each observation by
averaging the fast Fourier transform power spectra of continuous 128-s
intervals, with a Nyquist frequency of 4096 Hz and a lowest frequency
of 7.8 mHz. Periods of dipping activity in the X-ray light curves, as reported by  
 \citet{Kuulkers10}, were not excluded from our timing analysis. The Poisson noise
spectrum is estimated with the analytical function of \cite{Zhang95}
and subtracted from this average power spectrum. The resulting power
spectrum is expressed in source fractional rms normalization
\citep{Vanderklis89}, using the average background rate during the
observation. Each power spectrum was fit with a multi-Lorentzian (1--5
components) function and (when necessary) a power law.

We performed a spectral analysis for only a few selected observations,
including those cases where HID location and timing properties were
inconclusive with respect to the spectral state of the source. Spectra
were extracted from the Standard 2 mode data of PCU2, using HEASOFT
v6.10. The spectra were background subtracted, corrected for dead
time, and a systematic error of 0.6\% was applied. Fits were made with
XSPEC v12.6.0 \citep{Arnaud96} between 3.0--40.0 keV, with
phenomenological models consisting of a simple accretion disk
component (diskbb), a (cut-off) power-law, and a Gaussian line (fixed
at 6.4 keV). Reported fluxes are in the 2--20 keV band and are
corrected for interstellar absorption, which was fixed to
$1.7\times10^{21}$ atoms\,cm$^{-2}$, the average value in the
direction of J1659 \citep{Kalberla05}.

\begin{figure}
\includegraphics[width=1\columnwidth, height=9cm,angle = -0]{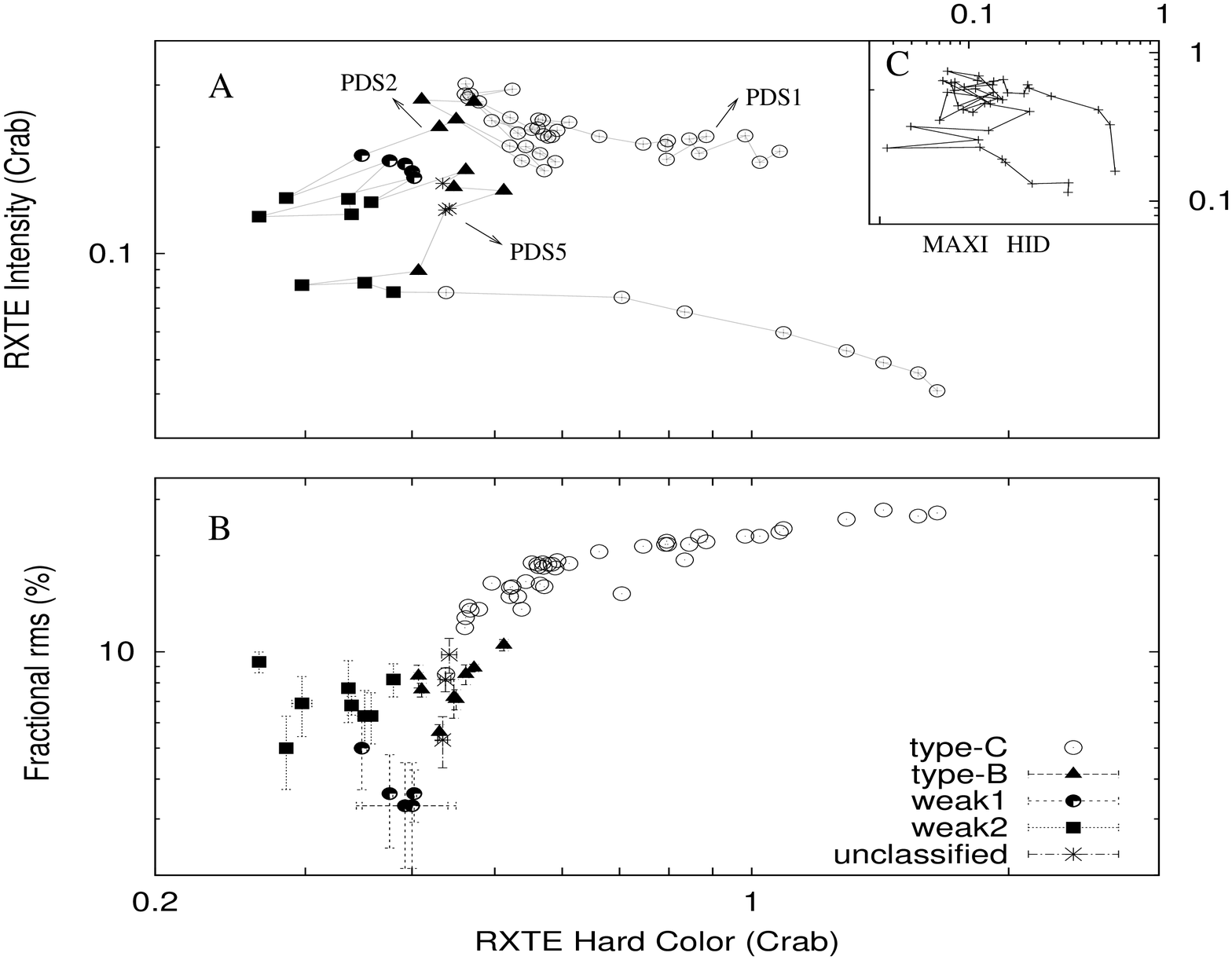}
\caption{With the {\it RXTE} data, we plot A) Hardness--intensity diagram (HID) and, B) 0.01-100 Hz fractional  rms amplitude vs. hard color. Each point represents one  observation, except for the MJD 55488.02 and 55490.11 observations, for which we use two points each (see text). Different power spectral types are indicated (symbols as in Figure \ref{fig:inhcrms}).  The locations of three of the power spectra shown in Figure \ref{pds} (PDS1, PD2, and PDS5) are indicated with arrows. With the {\it MAXI} data, C) HID, the intensity and hard color measured in 1.5--20.0 keV band and in 10.0--20.0/2.0--4.0 keV band, respectively.} \label{fig:hcinrms}
\end{figure}

\section{Results}\label{results}

\subsection{Light curves and color evolution}\label{sec:lc}

The top panel of Figure \ref{fig:inhcrms} shows the 1.5--20 keV light
curve from {\it MAXI}. It includes the early rise and late decay,
which were not covered by our {\it RXTE} data set (see below).  Panels B--E of Figure \ref{fig:inhcrms} show the {\it RXTE}
2--20 keV light curve, hardness curve, time evolution of fractional
rms amplitude (0.01--100 Hz), and QPO frequency evolution,
respectively. The outburst showed a relatively rapid rise; when it was
first observed with {\it RXTE}, three days after discovery, it already
had a flux of 6.6$\times$10$^{-9}$ ergs/cm$^{2}$/s. The rise
was followed by an extended plateau during which some irregular
intensity variations (time scale of a few days) were observed
(hereafter referred to as `flares'). During two observations in this
flaring phase (MJD 55488.02 and 55490.11) we observed several rapid
($\sim$20 s), almost step-function-like changes in the count rate from
the source. The changes were on the order of 30\%, lasted
$\sim$100--1500 s, and were not accompanied by changes
in the spectral hardness. Similar phenomena have been observed in,
e.g., GX 339-4 and XTE J1859+226 \citep{Miyamoto91,Casella04} and have
been referred to as `flip-flops'. For both observations the high and
low count rate levels were separated and treated as different
observations for the remainder of our analysis (and are plotted
separately in Figures \ref{fig:inhcrms}--\ref{fig:hcinrms}). The
flux reached its maximum during a flare on October 8 (MJD 55477)
at a value of 1$ \times$10$^{-8}$ ergs/cm$^{2}$/s.  A sudden radio
flux decrease was reported for that day \citep{Vanderhorst10a}.
From the intensity plateau the source evolved towards a slow decay,
after which it presumably returned to quiescence.  The flux during the
last {\it RXTE} observation was 1.4$\times$10$^{-9}$ ergs/cm$^2$/s.

\begin{figure*}
\center
\resizebox{0.75\columnwidth}{!}{\rotatebox{0}{\includegraphics{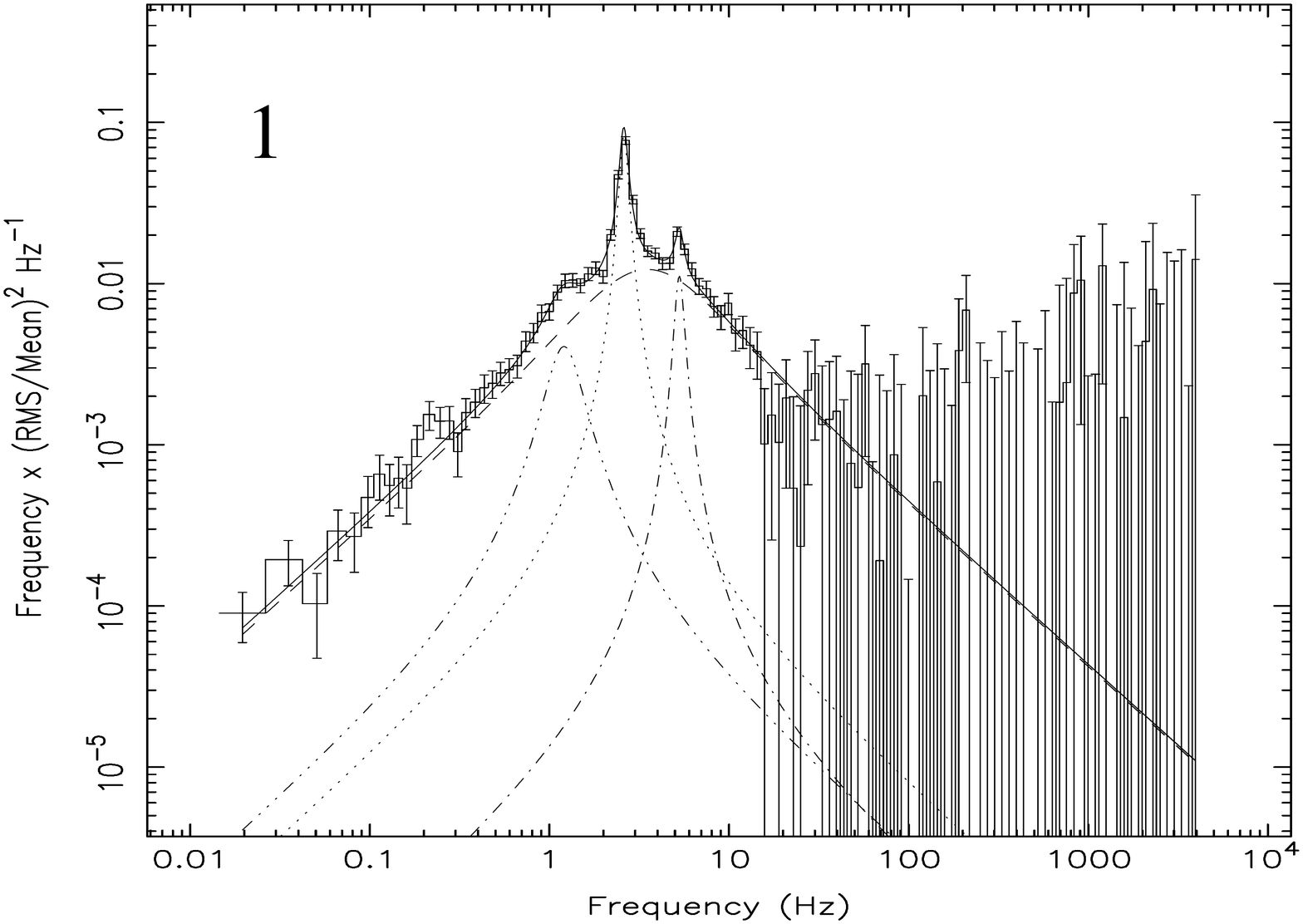}}}
\resizebox{0.75\columnwidth}{!}{\rotatebox{0}{\includegraphics{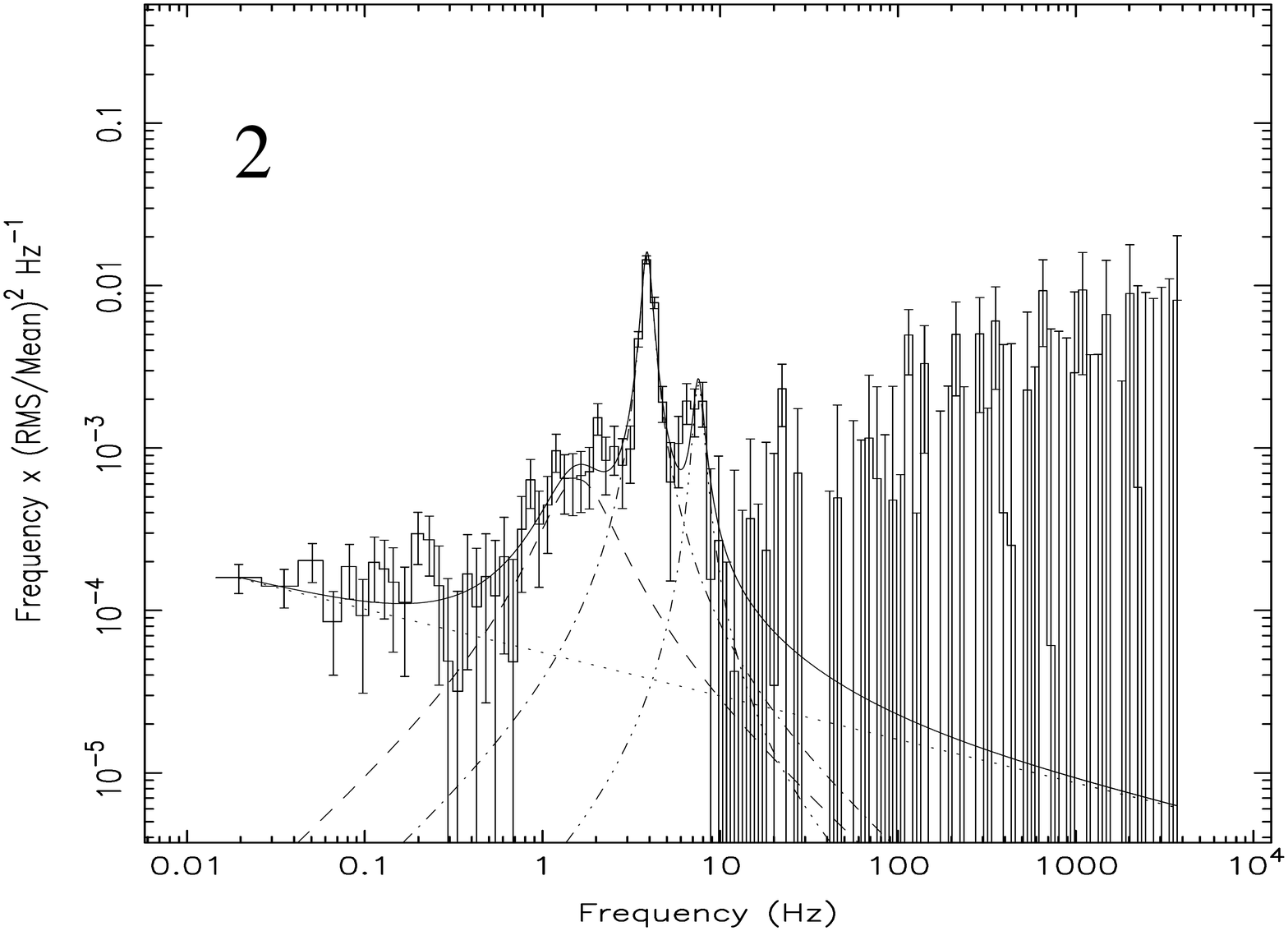}}}
\resizebox{0.75\columnwidth}{!}{\rotatebox{0}{\includegraphics{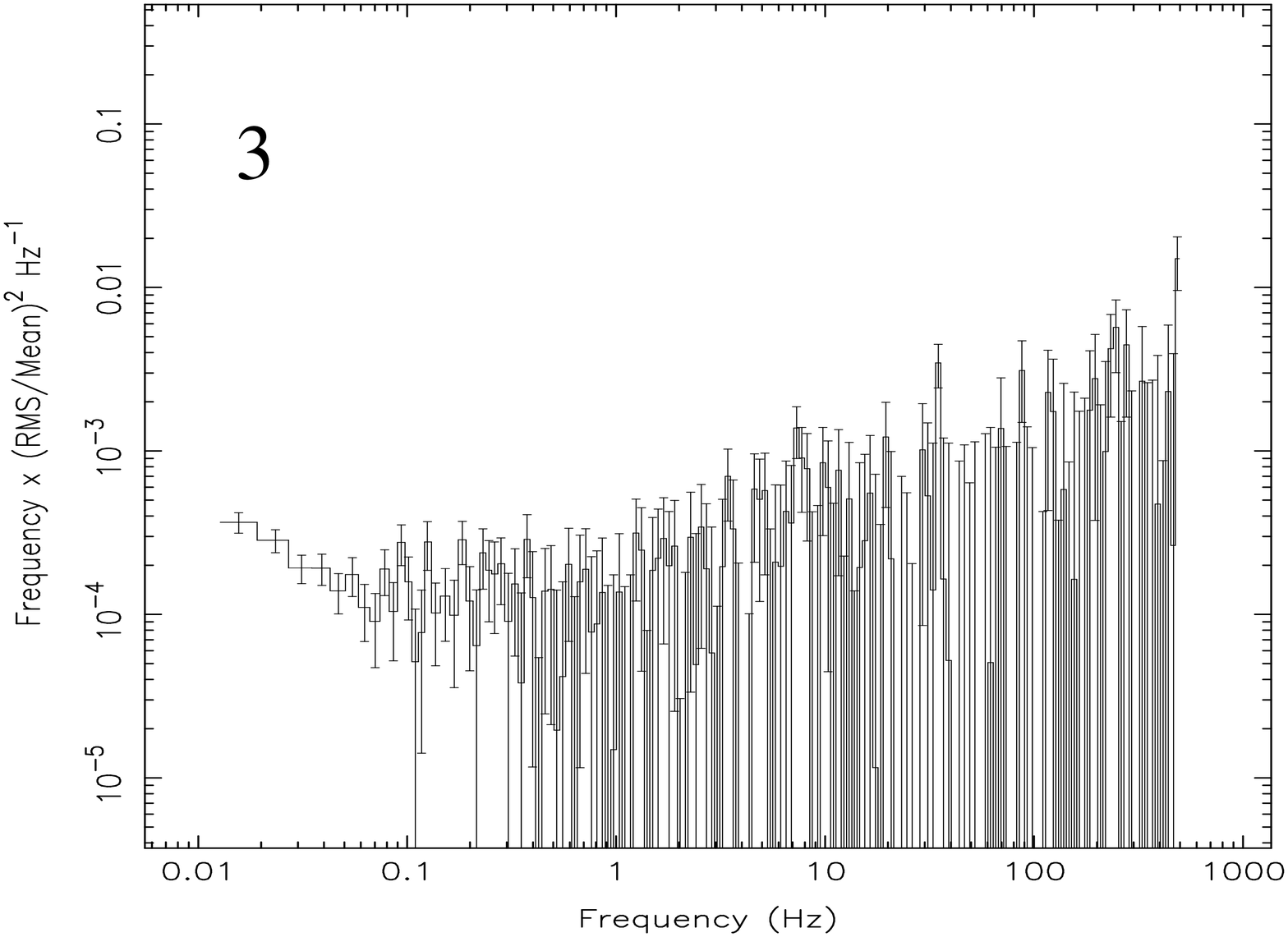}}}
\resizebox{0.75\columnwidth}{!}{\rotatebox{0}{\includegraphics{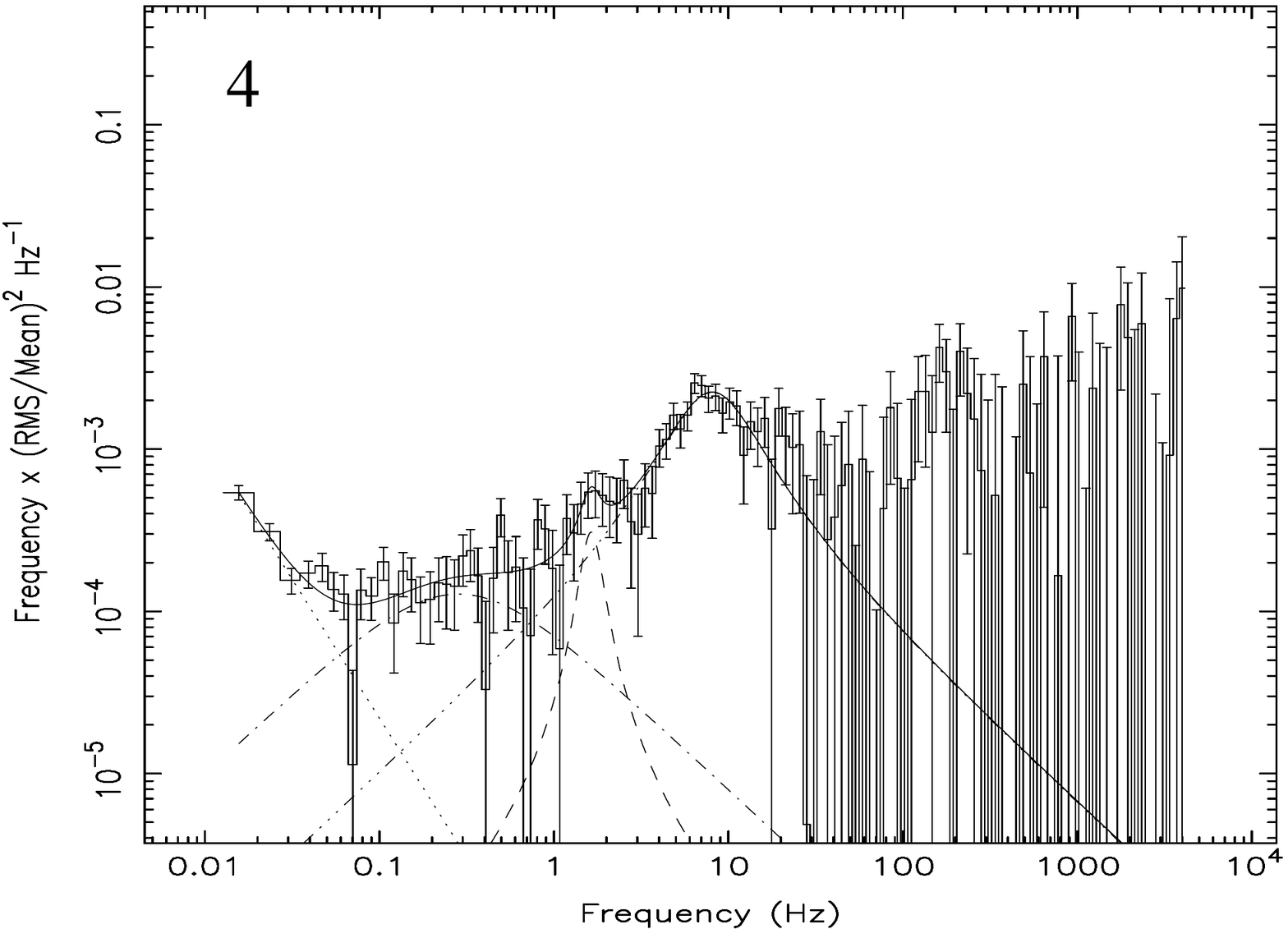}}}
\resizebox{0.75\columnwidth}{!}{\rotatebox{0}{\includegraphics{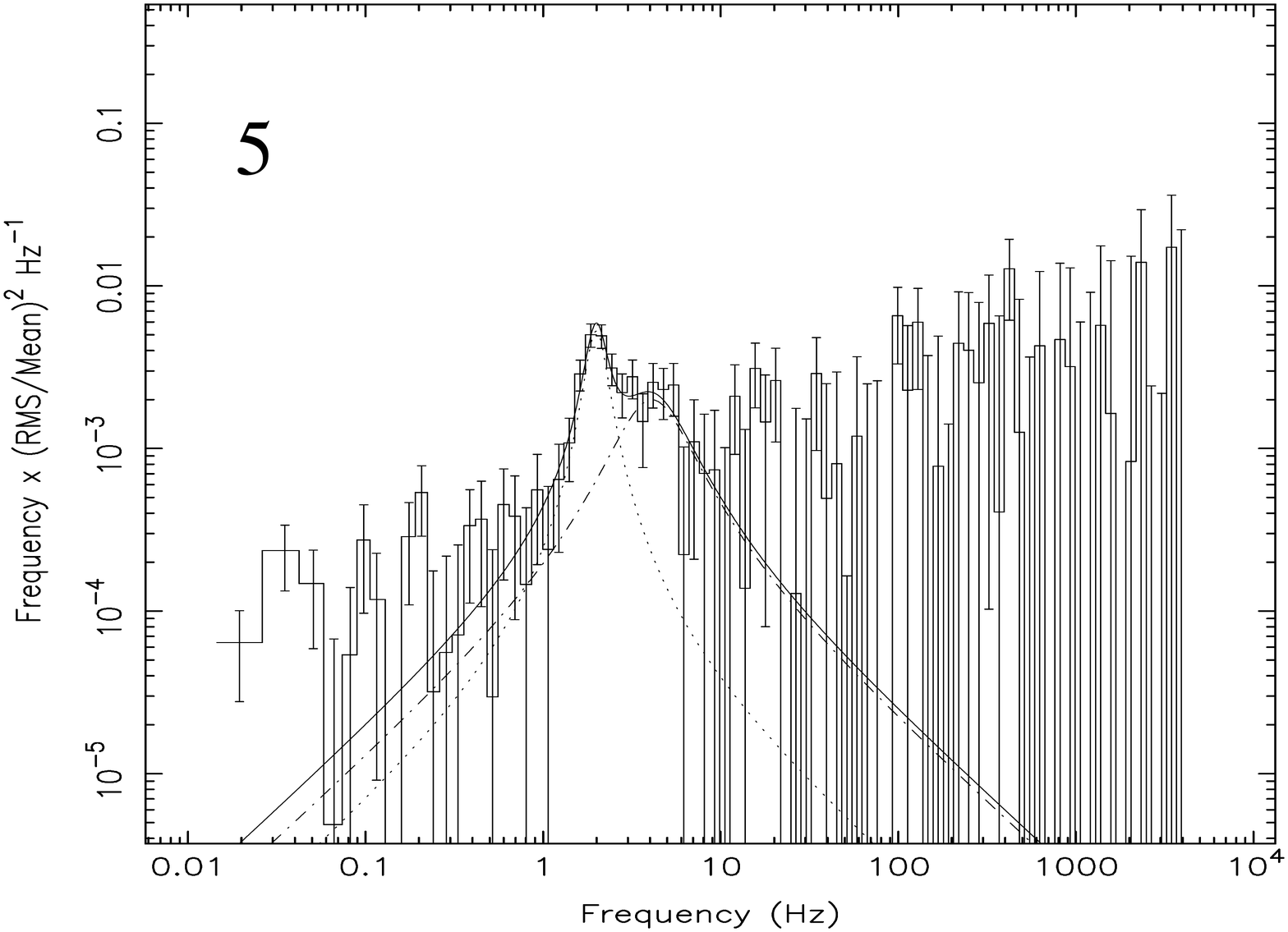}}}

\caption{Five representative power spectra, labeled PDS 1--5. PDS 1
  (ObsID: 95108-01-02-00) and PDS 2 (ObsID: 95118-01-01-01) are
  typical of type-C and type-B QPOs, respectively. PDS 3 is an average
  power spectra of the five observations classified as `weak1'. PDS 4 is the
  average power spectra of the nine observations classified as `weak2'. PDS 5 (ObsID: 95118-01-12-00) is the power spectrum of
  an observation which had an unclassified QPO.   The HID locations of PDS 1, 2 and 5 are indicated in Figure~\ref{fig:hcinrms}.}
\label{pds}
\end{figure*}

 Figure \ref{fig:hcinrms}A shows the HID. The source started in the upper-right corner and traversed its track in an anti-clockwise manner, ending in the lower-right corner. Hysteresis is observed in the spectral evolution, with the hard-to-soft spectral evolution occurring at a count rate $\sim$3 times higher than the soft-to-hard spectral evolution. While the spectral evolution during the early and late phases of the outburst was largely monotonic,  the source exhibited frequent back-and-forth motion in spectral hardness during the flaring phase of the outburst. 
 
Since {\it RXTE} did not observe J1659 during the first three days of its outburst, we used {\it MAXI} \citep{Matsuoka09} data to infer the spectral evolution during the early rise. We plot the {\it MAXI} HID in Figure \ref{fig:hcinrms}C. Overall, the path traced out is similar to that seen in the {\it RXTE} HID. The first part of the {\it MAXI} HID (i.e.\ the right vertical branch and subsequent left turn) was not covered by {\it RXTE}. A comparison with HIDs of other BHTs suggests that during the first three days of the outburst, J1659 evolved through parts of the LHS and HIMS.

\citet{Shaposhnikov10} reported that during the softest part of the outburst J1659 reached the HSS. We performed a spectral fit of the observation with the lowest spectra hardness (MJD 55489.26) to verify this. We obtain an excellent fit (reduced $\chi{^2}$=0.96, for 62 d.o.f.) with a 0.8 keV disk black-body and a power-law component with index 2.2 (plus a weak Gaussian at 6.4 keV).  The disk component contributed $\sim$60\% to the 2--20 keV flux. \citet{Belloni10b} does not provide a spectral definition of the HSS, but following the definition of the soft (or thermal dominant) state by \citet{Remillard06}, which uses 75\% as a lower limit, we conclude that the source did not reach the soft state during this outburst. This
behavior is similar to, e.g., the 2000 outburst of XTE J1550--564 \citep{Miller01}.

\subsection{Variability properties}

Figure \ref{fig:hcinrms}B shows the relation between hard color and broad-band rms. Such a diagram can be helpful in identifying different types of power spectra \citep{Belloni05,Fender09}. At high hard colors, hardness and rms are correlated, but at low hard colors ($<$0.5) there is considerable scatter. 

Based on the overall strength of the broad-band variability, the spectral hardness, and the shape of the power spectra (noise/QPOs) we divided our {\it RXTE} observations into five groups (see symbols in Figures  \ref{fig:inhcrms}-\ref{fig:hcinrms}). For the classification of the PDS and the QPOs, we made use of the classification schemes of \citet{Remillard02} and \citet{Casella05}. Representative power spectra are shown in Figure \ref{pds} and are referred to as PDS 1--5. 

\subsubsection{Group 1: type-C QPOs}

The Group 1 PDS have strong 0.01--100 Hz variability ($\sim$23\%), a strong QPO (varying between 1.6--8.5 Hz), at times with a harmonic and a sub-harmonic. They were observed during the spectrally hard parts of the outburst (hard color $>$ 0.45; see open circles in Figures \ref{fig:inhcrms} and  \ref{fig:hcinrms}). The frequency is strongly anti-correlated with spectral hardness (Fig.\ \ref{fig:inhcrms}E). The QPO frequencies and Q values, and the fractional rms amplitude in the 0.01--100 Hz frequency range are typical of PDS exhibiting type--C QPOs, as observed in the HIMS (no phase lags or energy dependence of the QPOs were measured). Hence, we identify these QPOs as type--C.

\subsubsection{Group 2: type-B QPOs}

The Group 2 PDS have 0.01--100 Hz variability $<$ 10\% rms and power-law noise as opposed to the strong variability and peaked noise of the Group 1 PDS. These PDS are observed in a fairly narrow range of spectral hardness, have QPOs with frequencies between 2.6 and 4.1 Hz, at times accompanied by a harmonic and a sub-harmonic, and cluster in a small patch of points in the hardness-rms diagram. In Figure \ref{rmsvsfreq}, we plot the 0.01--100 Hz fractional rms amplitude vs. frequency of these and the PDS with type--C QPOs, as well as the PDS with the unclassified QPOs (see below). What we observe in Figure \ref{rmsvsfreq} is very similar to the results of \citet{Casella04}, indicating that these are type--B QPOs, which classifies these observations as SIMS. Fractional rms amplitude anti-correlates with frequency for type--C QPO, while type--B QPOs do not follow such an anti-correlation. Note that after the detection of the first two type--B QPOs (at 3.9 Hz and 4.1 Hz), the spectrum temporarily hardened during the next two observations, and type--C QPOs (at 6.0 and 6.9 Hz) were observed, (see Fig.\ \ref{fig:inhcrms}) followed again by type--B QPOs (3.9 and 3.5 Hz). This indicates rapid SIMS/HIMS transitions \citep{Wijnands99} over a period of $\sim$6 days.

\subsubsection{Groups 3 \& 4: weak power}

In the soft part of the outburst most power spectra do not show clear
indications for QPOs. Such power spectra were observed during two time
intervals. During the first time interval (MJD 55486--55590) the
source intensity moved up and down on a daily basis --- this is also
the period during which the flip-flops were observed (see
\S\ref{sec:lc}). Inspection of the individual power spectra revealed
subtle differences between the low count rate and high count rate
observations. PDS3 is the averaged power spectrum of the five high
count rate level observations (weak1 in Figures \ref{fig:inhcrms} and
\ref{fig:hcinrms}); it is consistent with a single power-law with no
distinguishable features. The low-count rate observations show
indications for an additional broad bump around 7--8 Hz. Such a feature
is also seen for the observations in the second interval with weak	
power (MJD 55496.5--55498.5). PDS4 is the averaged power spectrum of
these nine observations (weak2 in Figures \ref{fig:inhcrms} and
\ref{fig:hcinrms}).
The PDS4-like power spectra occurred at slightly lower
hardness than the PDS3 ones. Power spectra similar to PDS3 and PDS4
were observed in GX 339-4 \citep{Belloni05} and were grouped together
with power spectra that show type-A (not observed here, see below) and
type-B QPO to define the SIMS. Therefore, also based on this
definition J1659 did not reach the HSS.

\subsubsection{Group 5: unclassified QPO}

\begin{figure}
\resizebox{1\columnwidth}{!}{\rotatebox{0}{\includegraphics{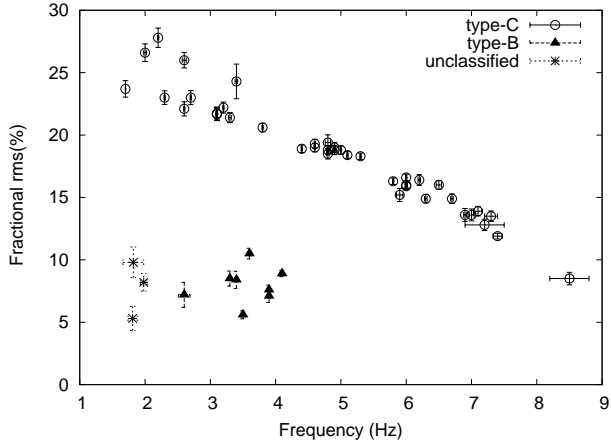}}}
\caption{0.01--100 Hz fractional rms amplitude vs. frequency of the strongest QPO in each power spectrum. The type--B and type--C QPOs form different branches, with the unclassified QPOs falling close to the type-B ones.}\label{rmsvsfreq}
\end{figure}

For three power spectra that showed somewhat broad QPOs (Q$\sim$1.6--3.5) we had difficulty assigning a classification --- they are marked as unclassified in Figures \ref{fig:inhcrms}--\ref{rmsvsfreq}. The observations were among the shortest (800 s) and the power spectra were of lower quality. However, the locations of these observations in Figure \ref{rmsvsfreq} (and also Fig.\ \ref{fig:hcinrms}B) suggest that these QPOs may be type--B QPOs with a lower frequency ($\sim$1.9 Hz) and also lower Q-value, which would classify these observations as SIMS.

We did not see indications for type--A QPOs in our observations. These
are typically accompanied by similarly shaped noise as type--B QPOs,
but show no strong harmonic content (as seen in PDS2). Also in the
rms-frequency plot they fall in a different location than type--B QPOs
\citep{Casella05}, which argues against the QPOs in the type PDS5-like
power spectra being of type--A.

\subsubsection{Search for High-frequency QPOs}

High-frequency QPOs in the the 50--450 Hz range have been reported for
a handful of BHTs \citep{Mcclintock06}. We inspected all individual
power spectra in the total 2--60 keV band, as well as in the 5.7--60
keV band, but no indications for such QPOs were found. We also
inspected the averaged power spectra of the different groups (with
groups 2 and 5 added together). Again, no significant QPOs were found,
with a 2.1$\sigma$ measurement of a peak at $\sim$160 Hz (Q$\sim$4,
$\sim$4\% rms) in the 2--60 keV power spectrum of PDS 4 constituting
our most significant detection.

\section{Discussion \& conclusions}

We analyzed all the {\it RXTE} observations of MAXI~J1659--152 taken
during its 2010 outburst. Spectral states were identified using the
HID in combination with the aperiodic variability properties. In the
HID J1659 traced out an anti-clockwise loop, similar to those
seen in black hole and neutron-star transients (see
\S\ref{intro}). The types of power spectral features we identified (in
particular type--B and type--C QPOs) and the way they evolve along the
HID track, indicates that J1659 is a black hole candidate.
The spectral decomposition of the softest spectrum (i.e.\ diskbb +
power-law, without the need for second thermal component;
\citealt{Lin07}), the radio loudness in the early part of the outburst
\citep{Paragi10}, the flip-flops seen in two observations, and
the sudden ($\sim$day) drop in the rms (i.e. PDS3) at low-hardness,
are all characteristic of this type of systems
\citep{Dunn10,Miyamoto91,Fender09}, strengthening our identification
of J1659 as a black hole candidate.

Based on spectral and variability properties we find that J1659
moved from a HIMS to a SIMS and back again (at lower intensity),
without reaching the HSS (or the soft or thermal dominant state as
defined by \citealt{Remillard06}). Our {\it RXTE} observations started
after the source had left the LHS (as indicated by {\it MAXI}
observations), and ended before it had returned to that state.

The observations in the softest part of the SIMS had weak variability and revealed two types of power spectra, one consistent with a power-law and one with an additional bump around 7--8 Hz. We note that similar groups of observations with weak variability were observed in the SIMS of GX 339--4 by \citet{Belloni05}. \citet{Fender09} found that radio flares and subsequent quenching of the radio flux often occur in a time interval of a few days before and after the time of sudden drops in the rapid X-ray variability (identified as a distinct zone in their rms-hardness diagrams). In J1659 a similar drop occurred on MJD 55481, a few days after the quenching of the radio flux as reported by \citet{Vanderhorst10a}, consistent with other BHTs.

Optical observations constraining the mass of the compact object are required to confirm the black hole nature of J1659.  From dipping episodes in the X-ray light curves, the orbital period period has been proposed to be 2.4--2.5 hour \citep{Kuulkers10,Belloni10a}. This makes the source very interesting as, if confirmed, this would make it the black hole binary with the shortest known orbital period.

\textbf{Acknowledgments} This research has made use of the {\it MAXI} data provided by RIKEN, JAXA and the {\it MAXI} team and has also made use of data obtained from the High Energy Astrophysics Science Archive Research Center (HEASARC), provided by NASA's Goddard Space Flight Center.  M. Linares acknowledges the support from NWO Rubicon Fellowship. P. Casella acknowledges funding via a EU Marie Curie Intra-European Fellowship under contract no. 2009-237722.

\end{document}